# Molecular Memory with Atomically-Smooth Graphene Contacts

Ahmad Umair[1], Tehseen Z. Raza[2] and Hassan Raza[1] *

[1] Department of Electrical and Computer Engineering, University of Iowa, Iowa City, Iowa 52242, USA

[2] Department of Physics and Astronomy, University of Iowa, Iowa City, Iowa 52242, USA

**Abstract:**

We report the use of bilayer graphene as an atomically-smooth contact for nanoscale devices. A two-terminal Bucky ball ($C_{60}$) based molecular memory is fabricated with bilayer graphene as a contact on the polycrystalline nickel electrode. Graphene provides an atomically-smooth covering over an otherwise rough metal surface. The use of graphene additionally prohibits the electromigration of nickel atoms into the $C_{60}$ layer. The devices exhibit a low-resistance state in the first sweep cycle and irreversibly switch to a high resistance state at 0.8–1.2 V bias. The reverse sweep has a hysteresis behavior as well. In the subsequent cycles, the devices retain the high-resistance state, thus making it write-once read-many memory (WORM). The ratio of current in low-resistance to high-resistance state is lying in 20-40 range for various devices with excellent retention characteristics. Control sample without the bilayer graphene shows random hysteresis and switching.

**Introduction:**

Reliable and efficient contacts are an important aspect of device design at the nanoscale. Historically, the contacts in the micron-scale devices have only been part of the overall device design for minimizing the contact resistance based on Schottky barrier height [1-3]. At the nanoscale, however, the influence of contacts on the transport channel is so important that an equal or often times even more effort is spent on the contact and interface design [4,5]. In various nanoscale devices, the contacts even dominate the transport characteristics [6,7]. While various novel contacts exist at the nanoscale with unique density of states, the simplest ones are the Ohmic contacts used to inject and extract the charge carriers. Mostly metallic contacts have been used as such contacts. However, in addition to the atomic roughness and grain boundaries, such contacts suffer from electromigration or filament formation, which may deteriorate the device characteristics and lead to reliability issues [8]. One of the grand challenges thus for the nanoscale design is to provide smooth and reliable contact to nanomaterials, being free from electromigration and any other non-ideal effects. In this paper, our objective is to explore graphene nanomembranes as a candidate for such contacts [9,10]. The use of graphene and boron nitride has been explored earlier for ultrathin circuitry [11].

In this work, we report the use of bilayer graphene (BLG) as an atomically-smooth contact in a molecular memory. Although various device structures based on graphene have been explored [12], our study is unique in the context of its use to improve reliability. The device schematic of the memory with BLG contact is shown in Figure 1. BLG is synthesized on a 300 nm evaporated Ni film by chemical vapor deposition (CVD). The detailed growth recipe is reported in the Method section. This layer prevents the electromigration of Ni atoms into the active material of the device. Furthermore, the use of BLG instead of monolayer or several-layer graphene is twofold. As compared to the monolayer, the probability of complete coverage with BLG is higher in the presence of defects and with increasing the number of layers, the transport properties of the device may be dominated by the multilayer graphene itself. Thus, BLG tends to provide an optimum solution. We then evaporate 100 nm $C_{60}$ film thermally, followed by evaporation of 5 nm Silicon dioxide ($SiO_2$) by electron-beam. $SiO_2$ film serves as the protective layer for the subsequent high-temperature metal evaporation. Finally, 90 nm thick top Cr electrode is deposited by using a shadow mask in an electron-beam evaporator. A control sample without BLG is also fabricated as shown in Figure 1.

A detailed characterization of the synthesized BLG has been reported earlier in Ref. [13]. The Raman spectrum of evaporated $C_{60}$ film on BLG is also shown in Figure 1. The dominant peaks are at 491 $cm^{-1}$, 1464 $cm^{-1}$ and 1596 $cm^{-1}$ wavenumbers, which confirms the coherence of $C_{60}$ molecular structure even after thermal evaporation [14,15].

**Results and Discussion:**

In Figure 2, we report the transport characteristics in the first and second sweep cycles for the device with BLG contact. The device starts in the low-resistance state and the voltage is increased in the forward direction till it irreversibly switches to high resistance state at about 0.9 V, as shown in Figure 2(a). After switching, the device withstands its high-resistance state, thus exhibiting hysteresis in the remaining cycle. We rule out the possibility of conductive filament formation (CFF) due to electromigration, since graphene has a breaking strength value of ~42 N/m and is impermeable even to helium atoms [16,17]. Moreover, in the CFF, current increases after switching, whereas opposite behavior is observed here. Apart from this, we find that the switching voltages for various devices lie in the 0.8 – 1.2 V bias range. This variation may be due to the amorphous and heterogeneous nature of the evaporated $SiO_2$ film [18].

The switching behavior for the second sweep cycle is shown in Figure 2 (b). The device remains in the high-resistance state without any hysteresis behavior. In the subsequent sweep cycles, the device sustains its high-resistance state, thus making it a possible write-once read-many (WORM) memory.

Apart from this, we report the retention characteristics in Figure 3, by using a read voltage pulse train of 0.4 V bias with 10 ms duration and 0.1% duty cycle. The mean value of current in the low-resistance state is 2.041 mA with a standard deviation of $0.973 \times 10^{-3}$. The device is then switched to the high-resistance state by applying a program voltage pulse of 1.2 V bias with 10 ms duration. The mean value of current is 89.29 µA with the standard deviation of 0.155. The current ratio of low-resistance to high-resistance state in this device is about 22.85 (which varied in 20-40 range for various devices). Besides the high retention time, the device also shows good endurance when continuous reading cycles with small pulse duration is applied. The retention characteristics are extrapolated to $10^4$ s, and a stable behavior is foreseen in both states of the device.

In Figure 4, we report the transport characteristics of the control sample without the BLG contact. The device shows random switching and hysteresis behavior. We observe that the current in this device is much higher as compared to the one with the BLG contact. Moreover, the current variation from device to device is considerable. We attribute this irregular behavior in our control sample to the atomically rough interface between Ni and $C_{60}$, as well as the electromigration of Ni atoms across $C_{60}$/Ni interface.

The switching mechanism in the reported WORM memory device with the BLG contact is not clearly understood yet, but covering Ni film with BLG prevents the electromigration of Ni atoms into $C_{60}$ film, thus stabilizing the device behavior. The transport characteristics do not show Ohmic or space-charge-limited conduction. Similar devices using $C_{60}$ molecules have been reported to have rewritable switching characteristics - quite different from our observation [19,20]. Moreover, multilayer graphene electrodes used in devices with PI:PCBM composite as active material have also been recently reported to have WORM memory behavior, whereas with the metallic electrodes rewritable switching characteristics has been reported [21]. Although the channel materials are different in the two experiments, the connection between the use of graphene and WORM features is noteworthy and needs to be explored further. Carbon nanotube [22] based contact have also been explored to eliminate electromigration, however, we believe graphene nanomembrane provides a better interface due to its 2D nature.

In conclusion, we have fabricated a molecular memory device with atomically-smooth BLG contacts. Covering Ni film with BLG shields the channel from metal surface irregularities and also prevents the electromigration of Ni atoms into the $C_{60}$ film. The device switches from a low-resistance to a high-resistance state, followed by hysteresis in the first sweep cycle. In the subsequent sweep cycles, the device remains in the high-resistance state and no hysteresis is observed, thus showing WORM memory behavior. The switching voltages vary in 0.8 – 1.2 V bias range for various devices with the high-resistance to low-resistance ratio vary in 20-40 range. The retention characteristics show good endurance under both low-resistance and high-resistance states up to $10^4$ s. In addition, replacing the top $Cr/SiO_2$ contact with BLG may further improve the characteristics, which we leave for future work.

**Method:**

We synthesized BLG on a 300 nm Ni film deposited on a 300 nm thermally grown oxide on Si substrate. Ni was deposited by using electron-beam evaporator (Angstrom Engineering) at 1Å/s rate under $< 7\times10^{-7}$ Torr chamber pressure. Ni pallets were placed in an Alumina boat (both supplied by International Advanced Materials) to avoid any contamination or residues. Prior to Ni evaporation, Si/SiO$_2$ substrate was cleaned with 10 min acetone, 10 min methanol, 10 min deionized (DI) water rinse, 20 min nanostrip (commercial Piranha substitute), followed by another 10 min DI water rinse. This sequence removes the impurities from the SiO$_2$ surface and provides better Ni adhesion. After Ni evaporation, the sample was further processed in UV ozone cleaner (Novascan PDS-UV), to remove any organic impurities on the Ni surface. The sample was then loaded into a home-made CVD furnace (Lindberg/Blue 1" diameter quartz tube) at room temperature under Ar ambient with 200 standard cubic centimeter (sccm) flow rate. After ramping the temperature to 1000 ˚C, the sample was annealed in H$_2$:Ar (65, 200 sccm) ambient for 10 min. BLG was then synthesized by flowing CH$_4$:Ar (23, 200 sccm) for 2 min, then moving the hot portion of the tube to the room temperature by ultra-fast cooling [13]. Research grade 5.0 (minimum purity 99.999%) process gasses supplied by Praxair Inc. were used. Thermo Scientific Nicolet Almega XR Raman spectrometer was used to characterize BLG with 532 nm laser (10 mW power) in the point scan mode with 15 s scan time and 4 scans per point. After BLG growth, 100 nm C$_{60}$ film was deposited by using thermal evaporator (Edwards Coating System E306A) at 1Å/s rate under $< 7\times10^{-7}$ Torr chamber pressure. The commercial C$_{60}$ powder was supplied by M.E.R Corporation. Raman spectroscopy was used to confirm the quality of evaporated C$_{60}$. A lower laser power of 2 mW with 5 s scan time and 4 scans per point is used to avoid sample heating. Next, the sample was loaded in the electron-beam evaporator, and 5 nm SiO$_2$ was evaporated, followed by 90 nm of Cr by using a shadow mask. The evaporation rates of SiO$_2$ and Cr were 0.2 Å/s and 1 Å/s, respectively, and the chamber pressure was $< 1\times10^{-7}$ Torr.

**Acknowledgment:**

We thank D. Norton, C. Coretsopoulos and J. Baltrusaitis for useful discussions. We acknowledge Micro-fabrication facility at the University of Iowa for evaporation, and Central Microscopy Facility at the University of Iowa for Raman spectroscopy. This work is supported by the MPSFP program of the VPR office at the University of Iowa.

**Figure Captions:**

Figure 1. Device schematics and characterization. Molecular memory with atomically-smooth bilayer graphene sandwiched between 300 nm Ni and 100 nm $C_{60}$ films, and control device without the bilayer graphene. Raman spectrum of evaporated $C_{60}$ film on the bilayer graphene is shown as well.

Figure 2. Transport characteristics in first and second sweep cycles. (a) During the first sweep cycle, the voltage is swept in the forward direction till it switches to high-resistance state. During the reverse sweep, the device remains in the high-resistance, and shows hysteresis. (b) The device remains in the high-resistance state during the second sweep cycle and no hysteresis or switching is observed.

Figure 3. Retention characteristics. The memory device shows a stable low-resistance state with for $10^3$ s. After switching to the high-resistance state by applying a 1.2 V write pulse of 10 ms duration, stable current is observed again. The dashed line is the interpolation to $10^4$ s.

Figure 4. Transport characteristics of the control device without the bilayer graphene. The voltage is first swept in the positive cycle, followed by a negative sweep cycle. A random switching and hysteresis behavior is observed.

Figure 1:

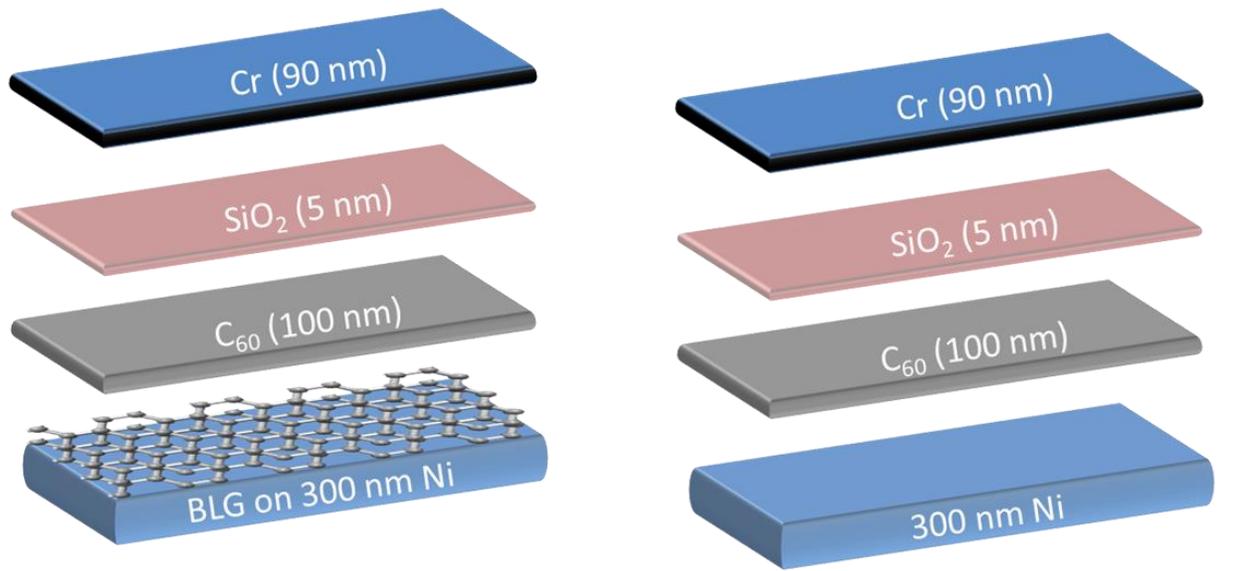

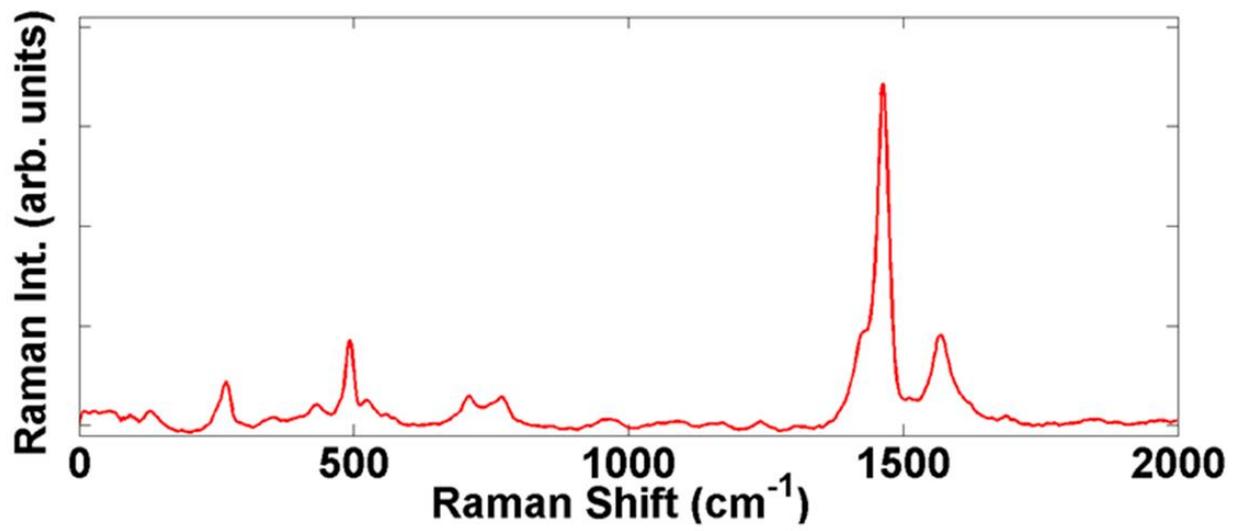

Figure 2:

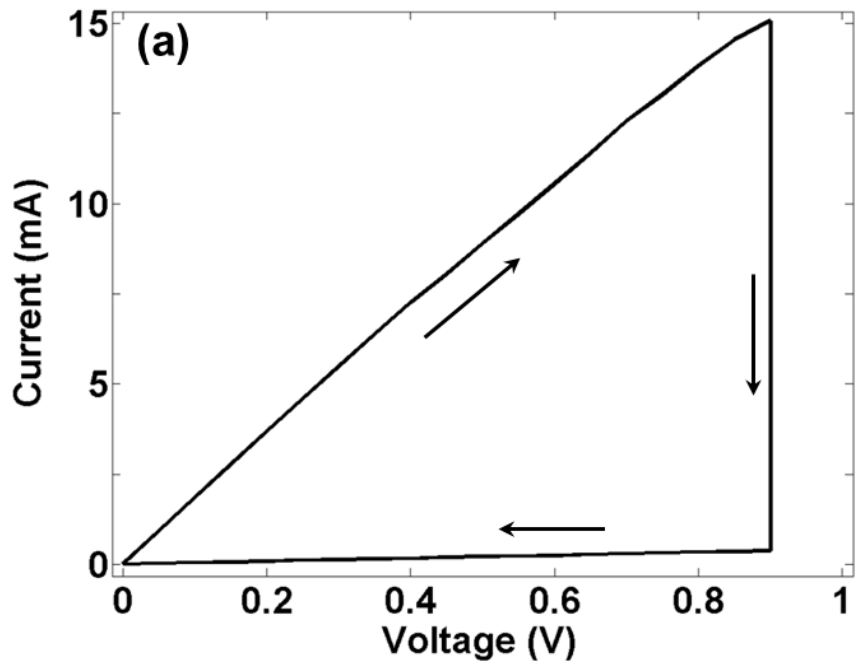

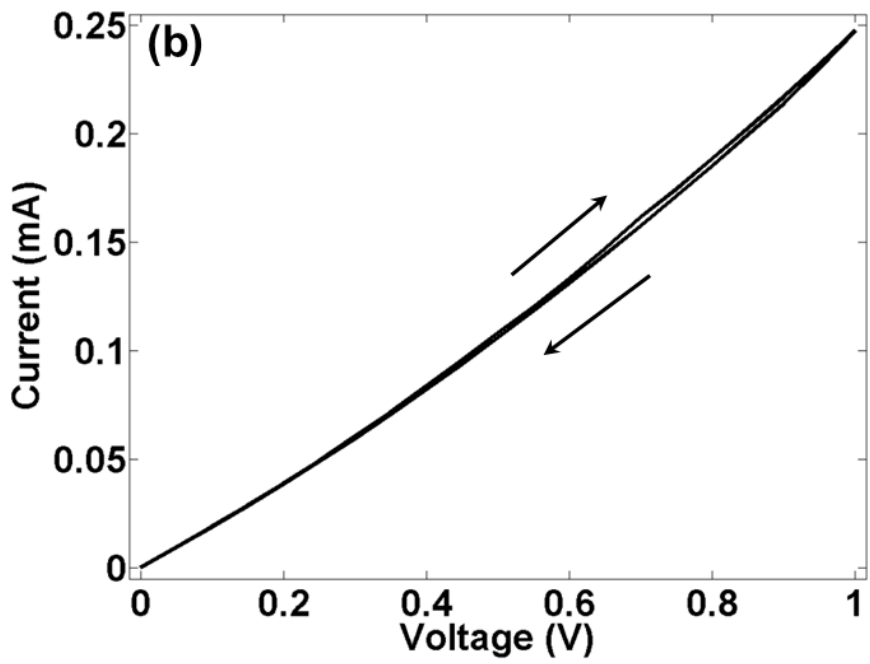

Figure 3:

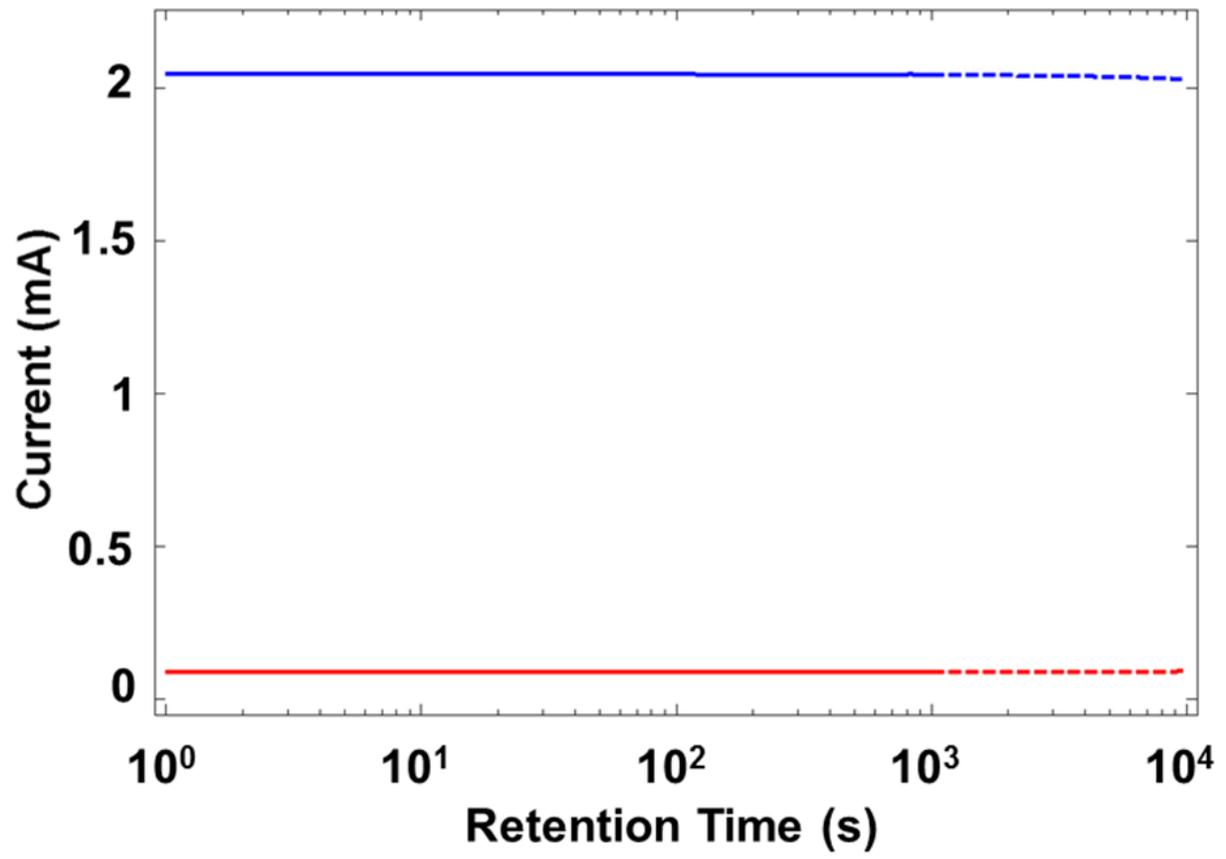

Figure 4:

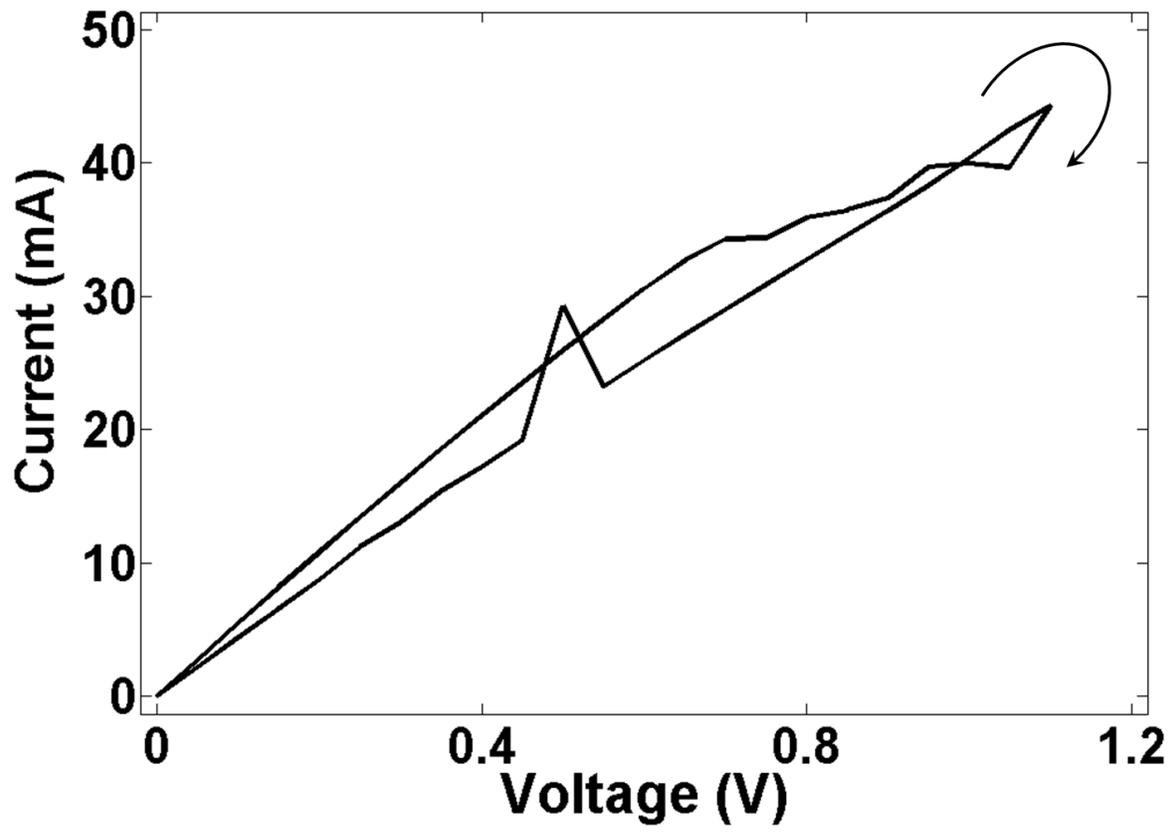